\documentclass[aps,pra,twocolumn,showpacs,floatfix,groupedaddress]{revtex4}

\usepackage{graphicx}
\usepackage{dcolumn}
\usepackage{bm}

\begin{document}

\title{Finite-temperature properties of quasi-2D Bose-Einstein condensates}

\author{Kwangsik Nho and D. P. Landau}

\affiliation{Center for Simulational Physics, University of Georgia, Athens,
Georgia 30602}

\date{\today}

\begin{abstract}
Using the finite-temperature path integral Monte Carlo method, 
we investigate dilute, trapped Bose gases in a quasi-two dimensional geometry.
The quantum particles have short-range, $s$-wave interactions described by
a hard-sphere potential whose core radius equals its corresponding scattering 
length. 
The effect of both the temperature and the interparticle interaction 
on the equilibrium 
properties such as the total energy, the density profile, and the superfluid
fraction is discussed. We compare our accurate results with 
both the semi-classical approximation and the exact results 
of an ideal Bose gas. 
Our results show that for repulsive interactions, 
(i) the minimum value of the aspect ratio, where the system starts to behave 
quasi-two dimensionally, increases as the two-body interaction strength 
increases, (ii) the superfluid fraction for a quasi-2D Bose gas is distinctly 
different from that for both a quasi-1D Bose gas and a true 3D system, i.e., 
the superfluid fraction for a quasi-2D Bose gas decreases faster than that 
for a quasi-1D system and a true 3D system with increasing temperature, and 
shows a stronger dependence on the interaction strength, 
(iii) the superfluid fraction for a quasi-2D Bose gas lies well below 
the values calculated from the semi-classical approximation, and 
(iv) the Kosterlitz-Thouless transition temperature decreases as the strength 
of the interaction increases.
\end{abstract}


\maketitle

The study of the effects of reduced dimensionality has attracted considerable
interest in condensed matter physics and statistical physics, for example, 
quantum films of superfluid $^{4}$He on surfaces, superfluid $^{4}$He clusters,
and superfluid $^{4}$He confined in restricted geometries. 
Recently, 
the experimental realization of Bose-Einstein condensation (BEC) in dilute, 
trapped, and supercooled atomic vapors~\cite{ADB} and
the additional ability to produce a low dimensional atomic gas trapped 
in an optical lattice have stimulated the investigation in a low dimensional 
system both theoretically~\cite{REV,THE1,THE2,THE3,THE4} and 
experimentally~\cite{EXP1,EXP2,EXP3,EXP4,EXP5,EXP6,EXP7,EXP8,EXP9,EXP10,EXP11,EXP12,EXP13,EXP14,EXP15,EXP16}. Dimensionally reduced systems
have very different properties from their three dimensional (3D) 
counterparts~\cite{MW,PCH} 
due to the enhanced importance of phase fluctuations. 
For example, in a spatially homogeneous infinite 
system a one-dimensional (1D) Bose gas does not exhibit BEC,
even at zero temperature; and in two dimensions (2D) BEC exists only at zero 
temperature. Nonetheless, a dilute two-dimensional Bose gas 
undergoes a superfluid phase transition at a finite critical temperature.
Below the Kosterlitz-Thouless transition temperature~\cite{BKT},
the gas is superfluid, and the superfluid phase is characterized by 
the presence of a quasicondensate, i.e., a condensate with only local phase 
coherence, since long wavelength phase fluctuations destroy long-range order.

Especially intriguing are two-dimensional 
Bose gases~\cite{THE5,THE6,THE7,THE8,THE9,THE10,THE11,THE12,REV2}, and
recently, experiments have entered regimes of BEC 
in 2D~\cite{EXP3,EXP4,EXP5,EXP6,EXP7,EXP8}. 
In a spatially homogeneous two-dimensional 
system Safonov {\it et al.}~\cite{EXP1}
observed the first experimental evidence for a quasicondensate in 2D atomic 
hydrogen gas adsorbed on a superfluid $^{4}$He surface. Furthermore, 
BEC in trapped gases is qualitatively different
from BEC in a homogeneous potential. 
Lower dimensional trapped Bose gases have 
been realized with strong quantum confinement in one or more directions~\cite{THE9}
(i) by gradually reducing atoms from a highly anisotropic trap to decrease 
the interaction energy~\cite{EXP3,EXP4}; 
(ii) by gradually increasing the trap anisotropy 
while keeping the number of atoms 
fixed~\cite{EXP5,EXP6}; and
(iii) by using the periodic potential of a one-dimensional optical 
lattice~\cite{EXP7,EXP8}. When the system is in a harmonic trap, 
the effect of the trap 
becomes more dramatic with lowering of dimensionality of the system;
the external trapping potential limits the size of the atomic gas,
and the density of states is modified.
As a result, Bagnato and Kleppner~\cite{BKr} showed that
an ideal 2D gas in a harmonic potential does exhibit a BEC phase at a finite
temperature. Petrov {\it et al.}~\cite{Petrov} showed 
that well below $T_{c}$ the equilibrium state is a true condensate,
whereas at intermediate temperatures a quasicondensate forms when local 
two-body interactions are included. 

Theoretically, the ground-state properties of a trapped Bose gas 
in two dimensions have been studied recently using the Gross-Pitaevskii 
mean-field theory~\cite{THH,TJS}, the leading quantum corrections to the 
Gross-Pitaevskii equation~\cite{THE6}, and 
a variational model based on a Gaussian-parabolic 
trial wave function~\cite{THE9}. 
For the finite-temperature properties of a dilute Bose gas confined in a 
harmonic trap, the semiclassical approximation~\cite{SSi}, 
a microscopic mean field 
theory that includes both density and phase fluctuations of 
the Bose gas~\cite{THE4},
the Hartre-Fock-Bogoliubov formalism~\cite{THE5,THE7,THE8}, and 
the scaling structure within diagrammatic perturbation theory~\cite{THE12} 
have been used.

However, many properties have yet to be investigated both experimentally 
and theoretically. Needless to say, a more thorough theoretical understanding 
is clearly desirable. At finite temperatures among the main problems to be 
answered are the density profile as a function of temperature and the strength 
of the interparticle interaction, under what conditions the confinement gives 
a system with dimensionality 2, the superfluidity, and the effect of 
interactions on the critical temperature. In addition, as the role of 
correlations and of quantum fluctuations is enhanced in quasi-2D geometries,
an appropriate theoretical description requires the more accurate 
many-body approach
beyond the mean-field approximation for a more systematic investigation 
in a wider range of temperature and values of the scattering length.

In this paper we use a finite-temperature path-integral 
Monte Carlo (PIMC) method~\cite{Ceperley} to investigate the crossover
from 3D to 2D for $N$ particles, where the interparticle interaction 
is a purely repulsive hard-sphere potential of radius $a_{s}$, the $s$-wave 
scattering length.
PIMC allows one to calculate accurate quantum mechanical expectation values
of many-body system, the only input being the many-body potential.
In particular, we demonstrate the influence of temperature $T$ and 
interparticle interaction $a_{s}$ on the equilibrium properties of 
ultracold atomic gases to 
calculate their energetics and structural properties in 
a highly anisotropic trap. Furthermore, we compare our accurate PIMC results 
with both the semi-classical approximation and the exact values 
for an ideal Bose gas, and we provide detailed predictions for comparison with 
future experiments.

We consider the following Hamiltonian for a system of $N$ hard spheres
\begin{eqnarray}
H &=& H_{0} + \sum_{i>j}^{N}v(r_{ij})\\
H_{0} &=& -\frac{\hbar^{2}}{2m}\sum_{i=1}^{N} \nabla_{i}^{2} + \frac{1}{2}m
\sum_{i=1}^{N}
(\omega^{2}_{x}x_{i}^{2}+\omega^{2}_{y}y_{i}^{2}+\omega^{2}_{z}z_{i}^{2}),
\nonumber
\end{eqnarray}
where $H_{0}$ is the Hamiltonian for trapped ideal Bose gases and $v(r)$
is a two-body, symmetric hard-sphere potential defined by 
\[v(r) =  \left \{\begin{array}{ll}
                   +\infty  & \mbox{($r < a_{s}$)} \\
                   0        & \mbox{($r > a_{s}$).} 
                  \end{array}
          \right. \]
In order to investigate the crossover from 3D to 2D,
we consider Bose gases under a cylindrically harmonic confinement 
($\omega_{x} = \omega_{y} = \omega_{\rho}$ and $\omega_{z}$ = 
$\lambda \omega_{\rho}$), 
where $\lambda$ is the aspect ratio. Here we shall always be concerned with 
the properties of trapped Bose gases at finite temperatures 
with the number of particles $N$,
the scattering length $a_{s}$, of the two-body interaction potential, and two 
characteristic lengths $a_{z}=\sqrt{\hbar/m\omega_{z}}$ and 
$a_{\rho}=\sqrt{\hbar/m\omega_{\rho}}$, describing the oscillation lengths 
in the transverse and the longitudinal directions, respectively.
We gradually vary the axial trap frequency to increase the trap 
anisotropy and to enter the quasi-two-dimensional regime, 
where the particles obey 2D statistics but 
interact in the same way as in a three-dimensional system.
As a result, the gas is confined by an extreme, pancake-shaped potential 
and has quasi-two-dimensional properties.
Along the tightly confined axial direction
the motion is frozen out, so that the condensate is in the harmonic 
oscillator ground state of an ideal gas, and the condensate width equals 
the harmonic oscillator length. 
The tight confining direction is characterized by the frequency $\omega_{z}$
and two weak confining directions by $\omega_{\rho}$.

\begin{figure}[]
\begin{picture}(0,200)(0,0)
\put(-130,210){\includegraphics{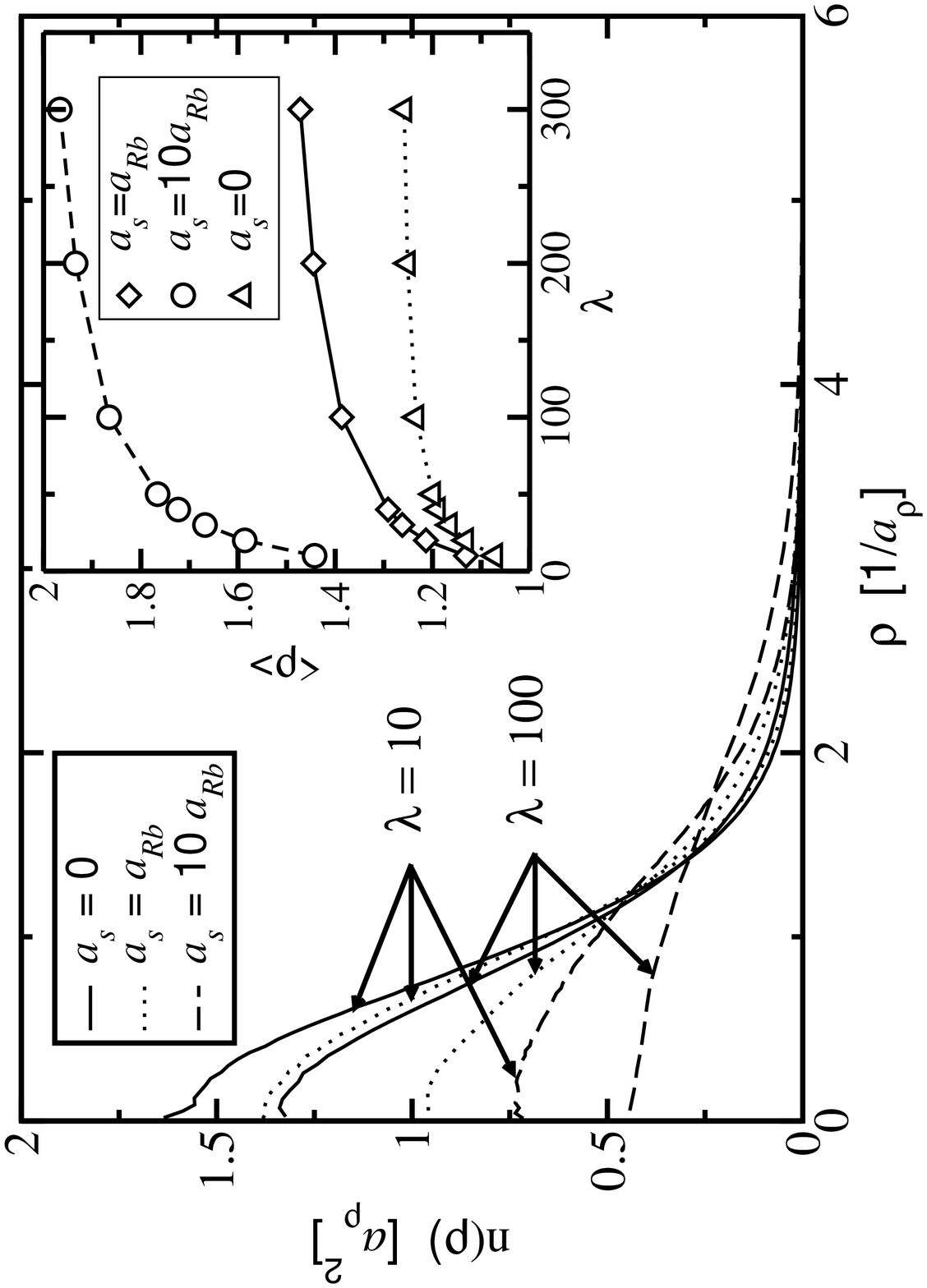}}
\end{picture}
\caption{The calculated density profiles $n(\rho)$, normalized 
such that $\int_{0}^{\infty} n(\rho)\rho d\rho = 1$ and 
$\rho = \sqrt{x^{2}+y^{2}}$, at $T$ = 0.4 $T_{c}$ for two different aspect
ratios, i.e., $\lambda$ = 10 and $\lambda$ = 100.
We used three different scattering lengths $a_{s}$ = 0 (solid lines), 
$a_{s}$ = $a_{Rb}$ (dotted lines), and $a_{s}$ = 10$a_{Rb}$ (dashed lines).
The inset shows the expectation value of $\rho$ in unit of $a_{\rho}$
as a function of $\lambda$ for $a_{s}$ = 0 (triangles), $a_{s}$ = $a_{Rb}$ 
(diamonds), and $a_{s}$ = 10$a_{Rb}$ (circles). 
In all figures, when statistical errors cannot be seen on the scale of the 
figure, the error bars are smaller than the symbol sizes.
The expectation value of $\rho$ clearly depends on both the scattering length 
$a_{s}$ and the aspect ratio $\lambda$. The dependence on $a_{s}$ 
becomes increasingly large as $\lambda$ increases.}
\label{fig1}
\end{figure}

\begin{figure}[]
\begin{picture}(0,200)(0,0)
\put(-130,220){\includegraphics{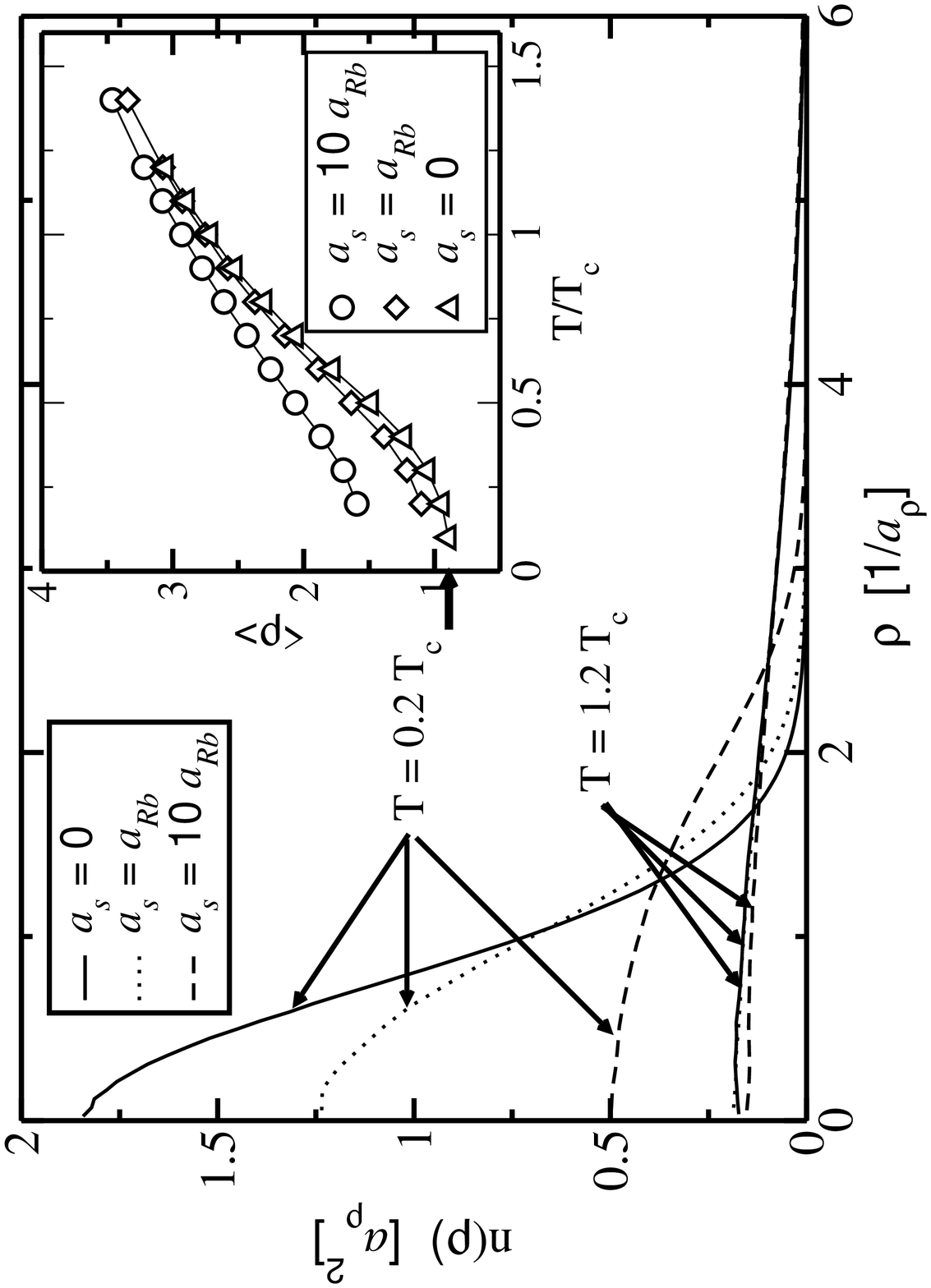}}
\end{picture}
\caption{
The calculated density profiles $n(\rho)$ at the aspect ratio $\lambda$ = 100
for two different temperatures, 
i.e., $T/T_{c}$ = 0.2 and $T/T_{c}$ = 1.2 using three different 
scattering lengths 
$a_{s}$ = 0, $a_{s}$ = $a_{Rb}$, and $a_{s}$ = 10$a_{Rb}$.
The inset shows the dependence of the expectation values of $\rho$
on the temperature $T/T_{c}$. The arrow on the y-axis indicates
the expectation value of $\rho$ at $T$ = 0, $<\rho>$ = 0.886 $a_{\rho}$. 
As the temperature increases, the density profiles $n(\rho)$ expand along 
the $\rho$ axis and the effect of the interaction strength decreases. 
}
\label{fig2}
\end{figure}

In this study, we typically use $N$=27 hard spheres 
because the permutation sampling 
in PIMC is efficient if $N$ = L$^3$, where L is an integer~\cite{PermS}, 
and the next largest number that satisfies this criterion would have required 
an excessive amount of cpu time. There is a finite-size effect in the density 
profile and the total energy; however, the superfluid fraction shows no 
dependence upon $N$ within the statistical error of the data. 
Experimentally, in an optical lattice, one can control and reduce 
the number of particle in each BEC~\cite{Markus,EXP8}.
Consequently, it will be 
possible to compare the effects predicted here for small numbers of atoms 
with experimental data directly.
We also use several trap aspect ratios 
1 $\leq$ $\lambda$ = $\omega_{z}/\omega_{\rho}$ 
$\leq$ 300 and a wide temperature range, i.e., 
0.1 $\leq$ $T/T_{c}$ $\leq$ 1.4,
wheret $T_{c}$ is the transition temperature for an ideal Bose gas of 
$N$ atoms in a trap (see, e.g., Eq.~(19) of Ref.~\cite{REV}).
Due to the finite-size corrections, 
the critical temperature $T_{c}$ depends on both 
the number of atoms $N$ and trapping frequencies, 
$\omega_{\rho}$ and $\omega_{z}$. 
For a repulsively interacting trapped Bose gas, $T_{c}$ is lowered compared 
to the trapped ideal Bose gas as a result of interaction, 
in contrast to an uniform 
Bose gas in the dilute range, where the critical temperature 
is increased above the ideal Bose gas value by interaction.
Here, the length unit is $a_{\rho}$, 
where $a_{\rho}$ = $\sqrt{\hbar/m\omega_{\rho}}$,
and energies are measured in units of $\hbar \omega_{\rho}$. 

In addition, we assume that 
the $s$-wave scattering length $a_{Rb}$ of $^{87}$Rb is 
100 times the Bohr radius, i.e., $a_{Rb}$ = 0.00433 $a_{\rho}$. 
PIMC procedure used here is based on methods described 
in our previous work~\cite{KL}.

We first discuss the density profiles of $N$ = 27 hard spheres 
at $T$ = 0.4 $T_{c}$ as a function of $\lambda$, 1 $\leq$ $\lambda$ $\leq$ 300.
Figure 1 shows the calculated density profiles
$n(\rho)$ as a function of $\rho$,  
normalized such that $\int_{0}^{\infty} n(\rho)\rho d\rho = 1$
and $\rho = \sqrt{x^{2}+y^{2}}$, for two different aspect ratios, i.e., 
$\lambda$ = 10 and $\lambda$ = 100.
To study the effect of the interaction strength, 
we used three different scattering lengths 
$a_{s}$ = 0 (solid lines), $a_{s}$ = $a_{Rb}$ 
(dotted lines), and $a_{s}$ = 10$a_{Rb}$ (dashed lines). 
Clearly, the total density
profile along the $\rho$ coordinate spreads out in the trap as both $a_{s}$
and $\lambda$ increases, i.e., 
the density at the center of the trap decreases with 
increasing $\lambda$ and $a_{s}$ as expected. 
In particular, the density profile $n(\rho)$ depends strongly 
on the interaction strength $a_{s}$ for $\lambda$ $\geq$ 10 and changes 
dramatically as a function of the aspect ratio $\lambda$. 
However, for $\lambda$ = 1 (a spherically symmetric harmonic trap), 
the density profile $n(\rho)$ for $a_{s}$ = $a_{Rb}$ approaches that 
of the non-interacting gas (see the inset of Fig.~1). 

The inset of Fig.~1 shows the expectation value of $\rho$ in unit of $a_{\rho}$
as a function of $\lambda$ for $a_{s}$ = 0 (triangles), $a_{s}$ = $a_{Rb}$ 
(diamonds), and $a_{s}$ = 10$a_{Rb}$ (circles). 
In all figures, when statistical errors 
cannot be seen on the scale of the figure, 
the error bars are smaller than the symbol sizes.
The expectation value of $\rho$ clearly depends on both the scattering length 
$a_{s}$ and the aspect ratio $\lambda$. The dependence on $a_{s}$ 
becomes increasingly large as $\lambda$ increases, which suggests that
the repulsion between the atoms spreads the atoms in the trap as a result of 
interaction.

Next, we further examine the condition to achieve a quasi-2D Bose gas using  
the density profiles $n(z)$ along the axial direction as a function of $z$,
normalized such that $\int_{-\infty}^{\infty} n(z) dz = 1$, and 
the expectation value of $\left|z\right|$. In the quasi-2D regime, 
along the $z$ direction the gas at finite temperatures 
has the characteristics of an ideal non-interacting gas at $T$ = 0.
Thus, the density profile $n(z)$ becomes identical to that of an ideal gas,
\begin{eqnarray}
n(z) = \frac{1}{\sqrt{\pi}a_{z}}\exp[-(z/a_{z})^{2}].
\label{nz}
\end{eqnarray}
From Eq.~(\ref{nz}), the expectation value of $\left|z\right|$ is 
0.564 $a_{z}$ for an ideal Bose gas at $T = 0$.

Our calculated expectation values of $\left|z\right|$ 
for $a_{s}$ = $a_{Rb}$ at $T$ = 0.4$T_{c}$ approach the ground state value 
($T = 0$) as the aspect ratio $\lambda$ increases.
Our calculated density profile $n(z)$ for $a_{s}$ = $a_{Rb}$ 
at $T$ = 0.4$T_{c}$ and that for an ideal gas (Eq.~(\ref{nz})) are
indistinguishable for $\lambda$ $\geq$ 40.
This suggests that the motion along the $z$ coordinate is indeed frozen out as 
$\lambda$ approaches large values since the energy of the axial excitations 
increases with increasing $\lambda$ and it is not easy for particles to go to 
the excited states for large values of $\lambda$. For repulsive interactions, 
the minimum value of $\lambda$, where the system starts to enter 
a quasi-2D regime, increases as the interaction strength increases.

To see the effect of the temperature $T$, we also calculated the density 
profiles $n(\rho)$ and $n(z)$ and the expectation values of $\rho$ and 
$\left|z\right|$ as a function of the scaled temperature $T/T_{c}$ for 
the aspect ratio $\lambda$ = 100 and $N$ = 27.

Figure 2 shows the calculated density profiles $n(\rho)$ as a function of 
$\rho$, for two different temperatures, i.e., $T/T_{c}$ = 0.2 and 
$T/T_{c}$ = 1.2 using three different scattering lengths 
$a_{s}$ = 0, $a_{s}$ = $a_{Rb}$,
and $a_{s}$ = 10$a_{Rb}$ (using the same symbols as in Fig.~1).
As the temperature increases, the density profiles $n(\rho)$ expand along 
the $\rho$ axis and the effect of the interaction decreases. 

The inset of Fig.~2 shows the dependence of the expectation values of $\rho$
on the temperature $T/T_{c}$. At very low temperatures,
our calculated expectation values of $\rho$ for $a_{s}$ = 0 approach 
the ground state value, i.e., $\left<\rho\right>$ = 0.886 $a_{\rho}$
(see the arrow on the $y$-axis).
As the temperature increases, the expectation values of $\rho$ increase, 
but the effect of the interaction decreases. 
The expectation values of $\rho$ for $a_{s}$ = 0 and $a_{s}$ = $a_{Rb}$ are 
the same within errorbars at high temperatures. 
However, the expectation values of $\rho$
for $a_{s}$ = 10$a_{Rb}$ are still larger than those for $a_{s}$ = 0 even at
high temperatures.

In contrast to the expectation value of $\rho$, the expectation 
value of $\left|z\right|$ for $\lambda$ = 100 
is nearly constant for $T$ $\leq$ 1.4$T_{c}$, 
the whole temperature range used here, and is the same as that for the ground 
state of an ideal Bose gas. Thus, the motion in the axial direction 
is largely frozen out and the system behaves quasi-two dimensionally 
at those temperatures. This demonstrates clearly that for the axial axis 
the system at finite temperatures behaves like an ideal gas at $T$ = 0.

\begin{figure}[]
\begin{picture}(0,200)(0,0)
\put(-130,200){\includegraphics{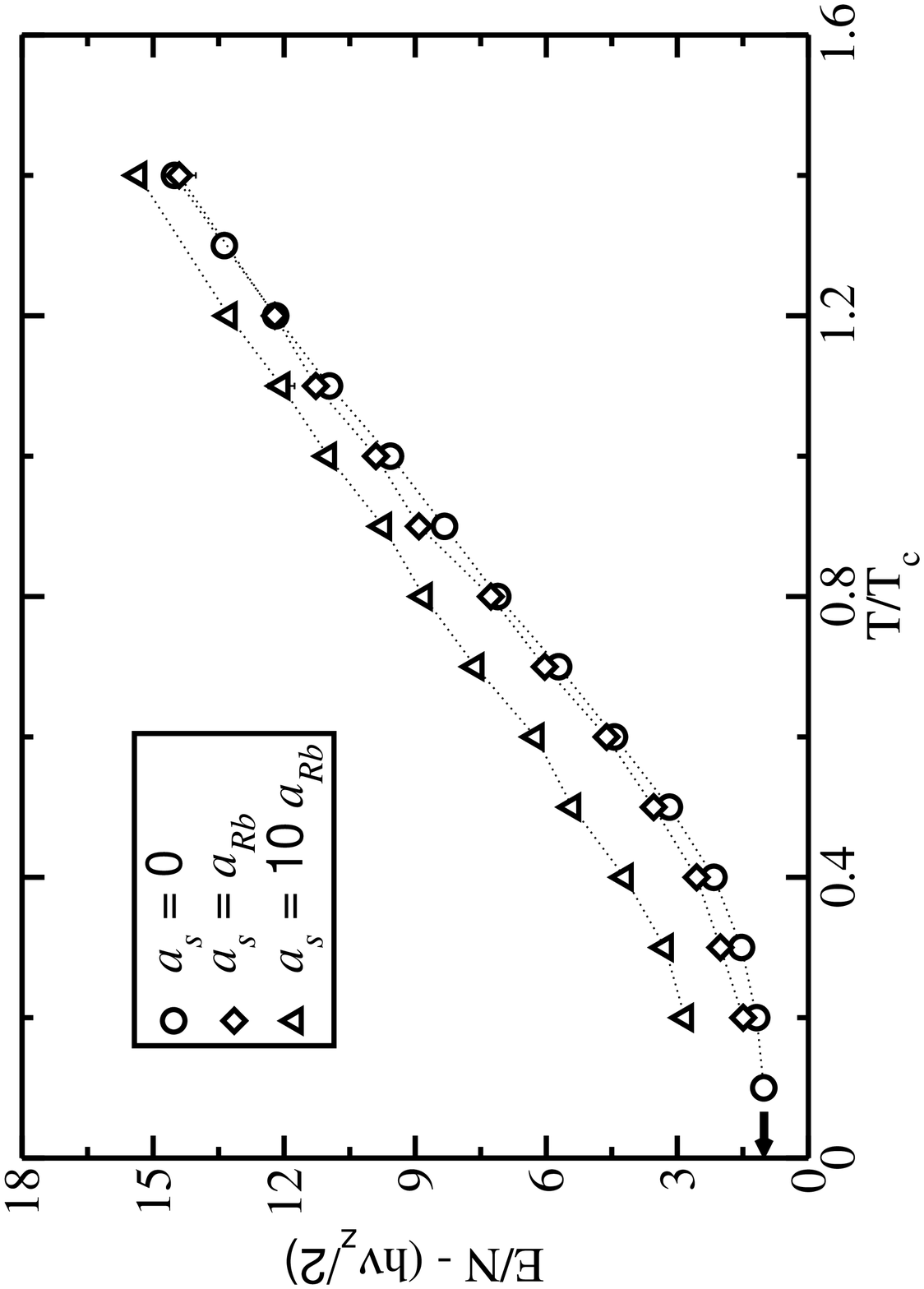}}
\end{picture}
\caption{
The calculated PIMC total energies per particle $E/N$ 
in units of $\hbar\omega_{\rho}$ as a function of the scaled temperature 
$T/T_{c}$ for $N$ = 27 and $\lambda$ = 100. We use three different interaction 
strengths $a_{s}$ = 0, $a_{s}$ = $a_{Rb}$,
and $a_{s}$ = 10$a_{Rb}$.
In the figure, we subtract $E_{0}^{z}$ = $\hbar \omega_{z}$/2 = 50 
$\hbar\omega_{\rho}$, the ground energy of an ideal Bose gas in the tight 
confinement direction, from our PIMC total energy.
The arrow on the y-axis indicates the ideal gas energy per particle 
in the $\rho$ direction, $\hbar\omega_{\rho}$, at $T$ = 0.
}
\label{fig3}
\end{figure}

\begin{figure}[]
\begin{picture}(0,200)(0,0)
\put(-130,200){\includegraphics{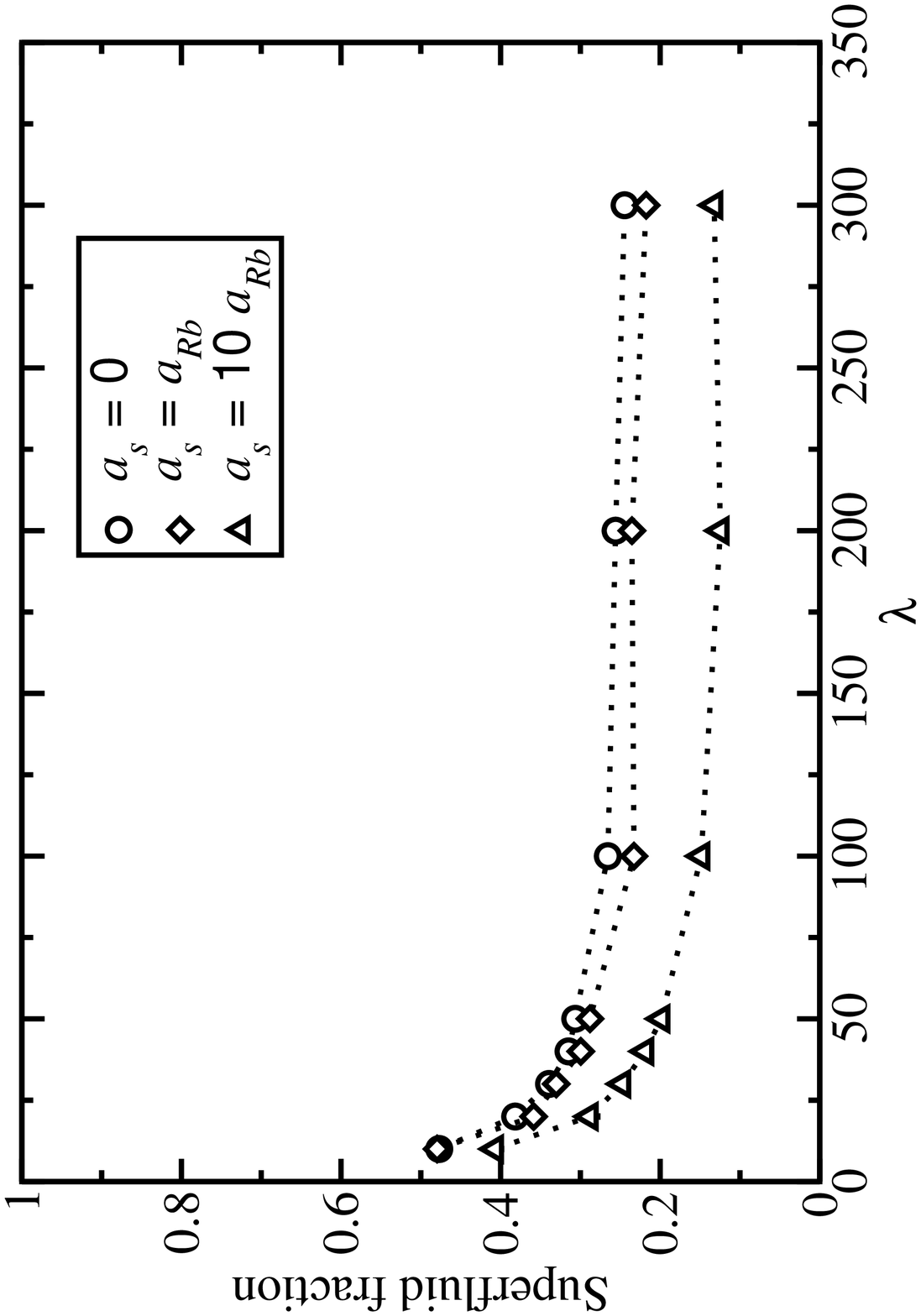}}
\end{picture}
\caption{
The superfluid fraction along the axis of rotation $z$ for $T/T_{c}$ = 0.4 and
$N$ =27 as a function of the aspect ratio $\lambda$ calculated 
using the PIMC method.
In this calculation we have used three different 
interaction strengths $a_{s}$ = 0, $a_{s}$ = $a_{Rb}$, and 
$a_{s}$ = 10$a_{Rb}$.
At the aspect ratios $\lambda <$ 100, 
the superfluid fraction decreases as the aspect ratio 
increases. However, the superfluid fractions do not depend on the aspect ratio,
to within error bars, for aspect ratios $\lambda$ $\geq$ 100. 
}
\label{fig4}
\end{figure}

Figure 3 shows our calculated PIMC total energies per particle $E/N$ 
in units of $\hbar\omega_{\rho}$ 
as a function of the scaled temperature $T/T_{c}$ for $N$ = 27 and 
$\lambda$ = 100. We used three different interaction 
strengths $a_{s}$ = 0, $a_{s}$ = $a_{Rb}$,
and $a_{s}$ = 10$a_{Rb}$.
For large anisotropies and small atom numbers, the total energy per particle
tends towards $\hbar \omega_{z}$/2, the ground energy of an ideal Bose gas 
in the tight confinement direction. As shown above, the motion along the $z$
direction is indeed frozen out at $\lambda$ = 100 for all temperature range 
used in this calculation 0.1 $\leq T/T_{c} \leq$ 1.4, i.e., the total energy 
per particle in the tight confinement direction equals that for an ideal 
Bose gas in one dimension, $\hbar \omega_{z}$/2 = 50 
$\hbar\omega_{\rho}$. 
In the figure, we subtracted  
$E_{0}^{z}$ = $\hbar \omega_{z}$/2 = 50 $\hbar\omega_{\rho}$ 
from our PIMC total energy per particle $E/N$ to interpret our results. 

Our calculated PIMC energies per 
particle for $a$ = 0  in the $\rho$ direction, $E/N$ - $E^{z}_{0}$, 
only approach the ideal gas 
energy per particle in the $\rho$ direction, $\hbar\omega_{\rho}$, with 
decreasing temperature as expected (see the arrow on the $y$-axis).
Above $T_{c}$ the total energies per particle for $a$ = 0 increase linearly 
with increasing temperature. However, below $T_{c}$  the total energy per 
particle decreases non-linearly with decreasing temperature, as expected
theoretically~\cite{REV}. 
The specific heat can be calculated by differentiating the total 
energy per particle with respect to the temperature to locate the critical 
temperature $T_{c}$, where the specific heat has a peak in plotting the values 
of the specific heat as a function of temperature.
In the present calculation, however, it is not easy to see a peak 
due to the finite size of our system and large error bars.
The scaled energies for $a$ = 10$a_{Rb}$ lie well above 
the non-interacting curve both below and above $T_{c}$.
The difference in the total energy per particle between for an ideal Bose gas 
and for a hard-sphere gas is visible at the low temperatures, and the clear 
difference increases as $T$ decreases because of the short-range structure 
of the hard-sphere potential, i.e., the comparison between the total energy 
for an ideal Bose gas and that for a hard-sphere gas shows that the scaled 
energies per particle clearly shows the effect of the interaction strength.

Finally, we calculated the superfluid fraction from PIMC data 
using the projected area~\cite{KL}, not using the winding number 
because we used open boundary conditions along all three axies. 
The superfluid fraction depends on the rotation axis for the anisotropic 
trap. In particular, we calculated the superfluid fraction with respect 
to the symmetry axis, 
i.e., the $z$ axis. Figure 4 shows the superfluid fraction as a function of 
the aspect ratio $\lambda$ calculated using the PIMC method.
In this calculation we used $T/T_{c}$ = 0.4, $N$ =27, and three different 
interaction strengths $a_{s}$ = 0, $a_{s}$ = $a_{Rb}$, and 
$a_{s}$ = 10$a_{Rb}$ 
(using the same symbols as in Fig.~3). For aspect ratios 
$\lambda <$ 100, the superfluid fraction decreases as the aspect ratio 
increases. However, the superfluid fractions do not depend on the aspect ratio 
within error bars for aspect ratios $\lambda$ $\geq$ 100. Figure 4 also 
shows that the superfluid fraction decreases gradually as both 
the two-body interaction strength and the aspect ratio increase.

\begin{figure}[]
\begin{picture}(0,200)(0,0)
\put(-130,220){\includegraphics{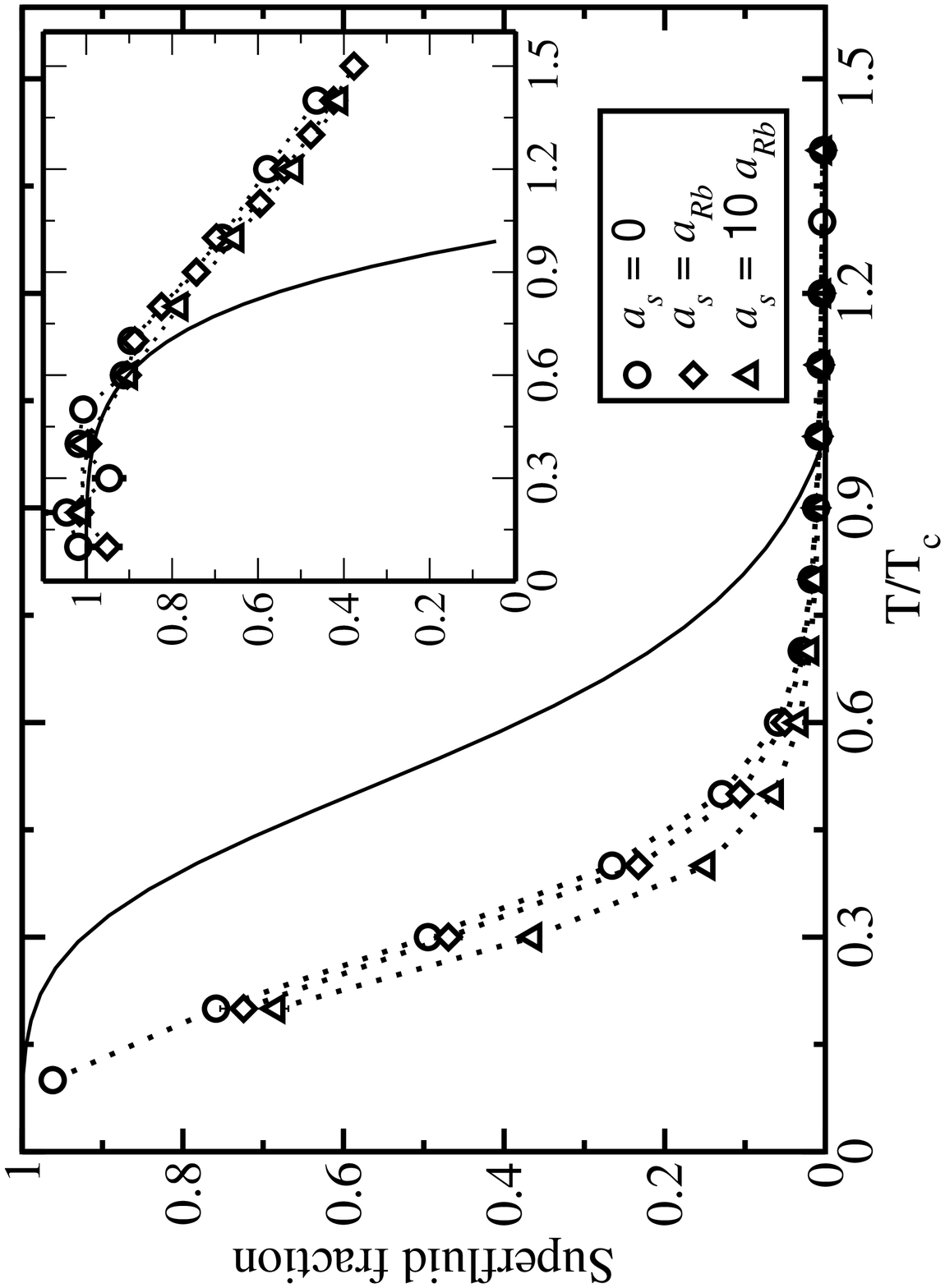}}
\end{picture}
\caption{
The calculated superfluid fractions along the $z$ axis at the aspect ratio 
$\lambda$ =100 and $N$ = 27 for three different interaction strengths
$a_{s}$ = 0 (circles), $a_{s}$ = $a_{Rb}$ (diamonds), and 
$a_{s}$ = 10$a_{Rb}$ 
(triangles) as a function of the scaled temperature $T/T_{c}$.
The inset shows the superfluid fraction along the $z$ axis 
for a quasi-1D Bose gas (using the same symbols as in Fig.~\ref{fig3}). 
Solid lines show the theoretical superfluid fraction
along the $z$ axis for an ideal Bose gas calculated from Eq.~(\ref{THEQ}).
}
\label{fig5}
\end{figure}

In Fig.~5 we present our calculated superfluid fractions for the aspect ratio 
$\lambda$ =100 as a function of the scaled temperature $T/T_{c}$. 
In the calculation, we used $N$= 27 and three different interacting strengths 
$a_{s}$ = 0, $a_{s}$ = $a_{Rb}$, 
and $a_{s}$ = 10$a_{Rb}$ (using the same symbols as in Fig.~\ref{fig3}). 
The superfluid fraction decreases from unity at very low temperatures and 
becomes zero above the critical temperature $T_{c}$ as the temperature 
increases.
The superfluid fraction has a small value at the same temperature for 
an interacting Bose gas compared to a non-interacting Bose gas, which shows
that the critical temperature for a repulsively interacting Bose gas is 
lowered compared to the ideal Bose gas as a result of interaction.
The inset of Fig.~5 shows the superfluid fraction for a quasi-1D Bose 
gas~\cite{Blume}.

The superfluid fraction for a quasi-1D Bose gas decreases much more 
slowly than that for a quasi-2D Bose gas with increasing temperature. 
For example, at the scaled temperature, $T/T_{c}$ = 1, 
the superfluid fraction for a quasi-1D system is significantly 
larger, about 0.65, compared to that for a quasi-2D gas, about 0.01. However,
the superfluid fraction for a quasi-2D gas shows a clear dependence
on the interaction strength whereas a quasi-1D gas shows
no noticeable dependence.

In order to compare our calculated PIMC data for the superfluid fraction
with the theoretical value of an ideal gas~\cite{SSi}, 
we calculated the theoretical 
superfluid fraction for an ideal Bose gas using Eq.~(14) of Ref.~\cite{SSi}.
The superfluid fraction $N_{s}/N$ is 1 - $\Theta/\Theta_{rig}$~\cite{Ceperley},
where $\Theta$ is the quantum mechanical moment of inertia and $\Theta_{rig}$ 
is the classical moment of inertia.
In our case, we used the $T_{c}$ with the first finite-size correction term
(see, e.g., Eq.~(19) of Ref.~\cite{REV}). For this, we modified 
Eq.~(14) of Ref.~\cite{SSi} by simply rewriting Eq.~(14) of Ref.~\cite{SSi}
using Eqs.~(8) and (9) of Ref.~\cite{SSi} to yield~\cite{Blume}

\begin{eqnarray}
\left(\frac{N_{s}}{N}\right)_{z} = \frac{1-(T/T_{c})^{3}}{1-(T/T_{c})^{3}
+1.801(T/T_{c})^{4}(k_{B}T_{c}/\hbar\omega_{\rho})}.
\label{THEQ}
\end{eqnarray}

\noindent
In order to derive Eq.~(14) of Ref.~\cite{SSi}, Stringari used 
the semiclassical approximation in the so-called macroscopic limit.
The solid lines in Fig.~5 and in the inset of Fig.~5 show the resulting 
approximate superfluid fraction. For a quasi-1D Bose gas, the theoretical 
values agree well with PIMC data below $T/T_{c}$ $\leq$ 0.6 and 
after $T/T_{c}$ = 0.6 the difference between the theoretical values 
and PIMC data gets larger as the scaled temperature increases. 
However, for a quasi-2D system, the values calculated from Eq.~(\ref{THEQ})
lie well above our calculated PIMC data below $T_{c}$.\\

In summary, using a finite-temperature path integral Monte Carlo 
technique we have investigated a trapped Bose gas in quasi-two dimensions
interacting via a hard sphere potential whose core radius equals 
its corresponding scattering length. In order to enter a quasi-2D regime, 
we have changed the axial trapping frequency $\omega_{z}$. 
We have presented and analyzed our accurate PIMC results such as the total 
energy, the density profile, and the superfluid fraction as a function 
of temperature $T/T_{c}$ at various values of the strength of the interaction
and the aspect ratio $\lambda$. We have compared our results with both 
the semi-classical approximation and the ideal Bose gas.
We find that for repulsive interactions, the minimum 
value of the aspect ratio $\lambda$, where the system start to behave 
quasi-two dimensionally, increases as the interaction strength increases.
In addition, for $N$ = 27 and $\lambda$ = 100, the motion along the axial 
direction is indeed frozen out for all temperature range used in this 
paper 0.1 $\leq T/T_{c} \leq$ 1.4 and the density profile $n(z)$ 
becomes identical to that of an ideal gas.
Specifically, we have calculated the superfluid fraction using the projected 
area and have compared the superfluid fraction for a quasi-2D system to that 
for a quasi-1D Bose gas.
The superfluid fraction for a quasi-2D Bose gas is distinctly different from 
that for both a quasi-1D Bose gas and a true 3D system, i.e., 
the superfluid fraction for a quasi-2D Bose gas decreases faster 
than that for a quasi-1D system and a true 3D system with increasing 
temperature, and shows a stronger dependence on the interaction strength.
In addition, the superfluid fraction for a quasi-2D Bose gas lie well 
below the values calculated from the semi-classical approximation and
the Kosterlitz-Thouless transition temperature decreases as the strength 
of the interaction increases.

\begin{acknowledgments}
We are greatly indebted to P.~Stancil and C.~D.~Fertig for their critical 
reading of the manuscript and illuminating comments. 
This work was partially supported by NASA grant No. NNC04GB24G.

\end{acknowledgments}


\begin{thebibliography}{lbl}
\bibitem{PermS}
The cyclic permutation of L atoms is accomplished via a random walk through 
permutation space. It is more efficient to use L $<$ $N$; however, 
it is not easy to determine how big L should be, 
although our experience suggests that L=$(N)^{1/3}$ works well.
\end{thebibliography}
\end{document}